\documentclass[10pt]{article}

\usepackage{amsmath}
\usepackage{amssymb}

\usepackage{graphicx}

\usepackage{cite}

\usepackage{color} 

\usepackage[labelfont=bf,labelsep=period,justification=raggedright]{caption}

\makeatletter
\renewcommand{\@biblabel}[1]{\quad#1.}
\makeatother

\date{}

\begin{document}

% Title must be 150 characters or less
\begin{flushleft}
{\Large
\textbf{Significant communities in large sparse networks}
}
% Insert Author names, affiliations and corresponding author email.
\\
Atieh Mirshahvalad$^{1,\ast}$, 
Johan Lindholm$^{2}$, 
Mattias Derl\'{e}n$^{3}$, 
Martin Rosvall$^{4}$
\\
\bf{1} Integrated Science Lab, Department of Physics, Ume{\aa} University, Ume{\aa}, Sweden 
\\
\bf{2} Department of Law, Ume{\aa} University, Ume{\aa}, Sweden
\\
\bf{3} Department of Law, Ume{\aa} University, Ume{\aa}, Sweden 
\\
\bf{4} Integrated Science Lab, Department of Physics, Ume{\aa} University, Ume{\aa}, Sweden 
\\
$\ast$ atieh.mirshahvalad@physics.umu.se

\end{flushleft}

% Please keep the abstract between 250 and 300 words
\section*{Abstract}
Researchers use community-detection algorithms to reveal large-scale organization in biological and social networks,
but community detection is useful only if the communities are significant and not a result of noisy data. 
To assess the statistical significance of the network communities, or the robustness of the detected structure, one approach is to perturb the network structure by removing links and measure how much the communities change. However, perturbing sparse networks is challenging because they are inherently sensitive; they shatter easily if links are removed. 
Here we propose a simple method to perturb sparse networks and assess the significance of their communities.
We generate resampled networks by adding extra links based on local information, then we aggregate the information from multiple resampled networks to find a coarse-grained description of significant clusters.
In addition to testing our method on benchmark networks,
we use our method on the sparse network of the European Court of Justice (ECJ) case law,
 to detect significant and insignificant areas of law. We use our significance analysis to draw a map of the ECJ case law network that reveals the relations between the areas of law.

% Please keep the Author Summary between 150 and 200 words
% Use first person. PLoS ONE authors please skip this step. 
% Author Summary not valid for PLoS ONE submissions.   
%\section*{Author Summary}

\section*{Introduction}

Network theory provides a good framework for studying systems composed of many interacting components. 
Recently, researchers have been interested in highlighting highly interconnected structures, communities, in biological and social networks \cite{girvan2002community,radicchi2004defining,Newman,danon2006effect,blondel2008fast,hastings2006community,rosvall2007information,palla2005uncovering},
because often communities correspond to behavioral or functional components.
For example, in social networks, communities can represent friendship groups; on the web, they can represent related pages on a specific topic; and in metabolic networks, they can represent cycles or other functional groupings. % Citations?
Here we show that communities can also capture disciplines of judgements in case law systems \cite{leicht2007large}.
However, similar to many real-world networks, the network of ECJ case law is sparse because of missing links.
% \add{Without an exact boundary between sparse and dense networks, we consider a network to be sparse if a clustering algorithm finds significantly more modules after a fraction of the links have been removed.}
The challenge in finding significant structures in sparse networks is twofold:
random noise directly propagates to the community results, and communities easily shatter because of missing links.
To find reliable communities in sparse networks with missing links, here we propose a simple method based on link prediction. First we show that our method performs well on benchmark networks. Then we apply our method to the ECJ case law network and generate a significance map of EU law.

Researchers use two main approaches to find statistically significant communities in networks: approaches based on explicit underlying null models in the clustering algorithms and approaches based on perturbation techniques. In the null-model approaches, communities are significant if the probability of finding them in a random network is lower than a given threshold \cite{Spirin2003,PhysRevE.81.046110,lancichinetti2011finding}. This is a solid approach when we are interested in how a network was formed. But when researchers are interested in highlighting functional aspects of an instantiated network, such as dynamics on a given network, they often use perturbation techniques \cite{gfeller2005finding,karrer2008robustness,PhysRevE.82.066106,rosvall2010mapping}. Taking this approach, researchers assume random noise in the data. When they perform the statistical analysis, they repeatedly perturb and cluster the data and then aggregate the results. Therefore, they can use any clustering algorithm and are not restricted to a particular null model. But for many sparse networks, the main source of error is not random noise in the data, but rather missing links with different effects on the clustering. For example, many clustering algorithms identify more clusters in sparse networks than in the corresponding networks without missing links \cite{PhysRevLett.101.078701,PhysRevE.80.056117}. 
Accordingly, we consider a network to be sparse if a clustering algorithm finds significantly more modules after a fraction of the links have been removed.
To take this shattering effect into account when we perform significance analysis on sparse networks with missing links, we introduce resampling based on link prediction.

To assess the significance of sparse networks with missing links, we combine perturbation techniques and link prediction. In practice, we resample sparse networks by completing triangles. For undirected networks, completing triangles corresponds to the simple and effective link prediction method called common neighbor \cite{Sociol1}.
With this approach, our aim is to add links that are missing because of insufficient data but avoid connecting nodes that factually are disconnected.
After explaining our approach in detail, first we show that we can recover shattered modules in benchmark networks as long as the mixing between modules is moderate and not too many links are deleted.
Then we apply the method to identify significant areas in the network of ECJ case law.
This network consists of more than 8,000 court cases connected by about 32,000 citations over the time period 1954--2010, clearly a sparse network.
We create a significance map and connect several insignificant clusters into complete areas of EU law.

% Results and Discussion can be combined.
\section*{Results and discussion}

\subsection* {Resampling based on completing triangles}
\begin{figure}[!ht]
\begin{center}
\includegraphics[width=0.5 \columnwidth]{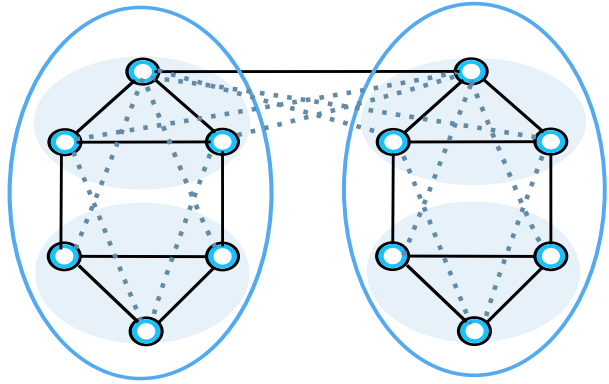}
\end{center}
\caption{
\textbf{Completing triangles followed by clustering aggregate shattered communities.} Dashed lines show different possibilities for completing triangles.}
\label{Schematic}
\end{figure}

To generate resamples of inherently sensitive sparse networks, %we could not use methods based on link deletion because they make the sparse network more shattered, However,
we need a method that efficiently adds extra links while preserving the core structure of the network.
That is, if we apply community detection algorithms for partitioning sparse networks with missing links, we will often find small shattered modules.
On the other hand, if we just add links randomly to prevent shattering, most likely we will connect nodes that should be disconnected because they are not directly related.
Accordingly, we note that the problem of aggregating shattered modules by adding links is similar to the problem of predicting missing links.
Missing links prediction methods operate by estimating the likelihood of a link between a pair of vertices
based on their similarity. To evaluate the similarity between vertices based on the structural properties of the network, indices like common neighbors \cite{Sociol1}, Jaccard coefficient \cite{Jaccard}, degree product, shortest paths, and hierarchical structure \cite{nature06830} have been proposed and used to predict future links on real data \cite{ASI20591}.
All similarity indices use specific assumptions about the positions of the missing links
that often make them complicated and computationally expensive to calculate.
But these assumptions might not reveal meaningful information in all real networks. %for example ---
To significantly analyze networks' communities by generating resampled networks, however,
we do not need to exactly predict missing links; we only need to add extra links in a non-destructive way so we can measure the robustness of the communities. 
Therefore, we perturb sparse networks with a simple and general method: triangle completion.
That is, we complete a fraction of open triangles that exist in the original data, see Figure \ref{Schematic}.
By adopting triangle completion, we assume that communities should have high density of triangles.
With this implicit null model, we can aggregate related shattered communities with a simple and general assumption about the network.
Triangles are the smallest unit of communities, and completing them strengthens local connections and the important core of the communities. As a result, shattered communities combine with each other and the community size grows. 

Figure \ref{Schematic} shows an example network in which black links indicate existing links in the  network and the four inner circles correspond to communities in the network. When we add links by completing the triangles (dashed lines), we aggregate the small communities into two big communities. Of course, by completing triangles we might add links between nodes that should not be connected. If more information is available about the network, other more sophisticated null models that may work better can be applied \cite{nature06830,ASI20591}. But as we show in the next section, the simple and general triangle completion method performs well on benchmark networks.

\subsection*{Benchmark networks}
\begin{figure}[!ht]
\begin{center}
\includegraphics[width=1 \columnwidth]{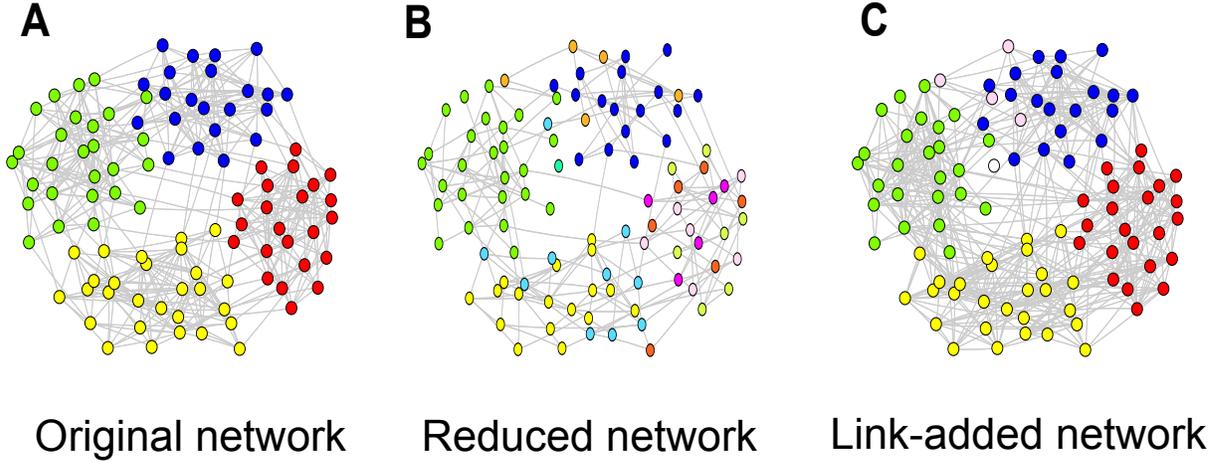}
\end{center}
\caption{
 \textbf{Triangle completion aggregates shattered modules.} Original network with 4 communities in A, removing links leads to small shattered communities in B, and completing triangles in the shattered network integrates small communities in C.}
 \label{Schem_bench}
\end{figure}

To validate our method, we tested triangle completion followed by clustering with the \emph{infomap} algorithm \cite{RosvallBergstrom08} on artificial networks with a built-in community structure. The benchmark graphs that we use resemble real-world network and was introduced by Lancichinetti \textit{et al.}\ \cite{PhysRevE.80.016118}. The benchmark networks have tunable exponents and we use exponent $-2$ for the degree distribution and exponent $-1$ for the community size distribution. Further, the mixing parameter $\mu$ determines the ratio between the external degree of a node with respect to its community and the total degree of that node. We use this framework to generate undirected networks with built-in community structures. 
Figure \ref{Schem_bench}A schematically shows a network with 100 nodes and four built-in communities. By removing 50\% of the links, communities fall apart and small modules are detected (Figure \ref{Schem_bench}B), But with triangle completion, related shattered modules are combined with each other (Figure \ref{Schem_bench}C).

To quantitatively show that triangle completion perturbs the network in a non-destructive way, we used normalized mutual information (NMI) % and preserving the coarse grain structure of communities, We should use a metric
to measure the similarity between the community structure of the original network and the community structure of the perturbed network \cite{danon2005,NJPhys.11.033015}. 

\begin{figure}[!ht]
\begin{center}
\includegraphics[width=0.7 \columnwidth]{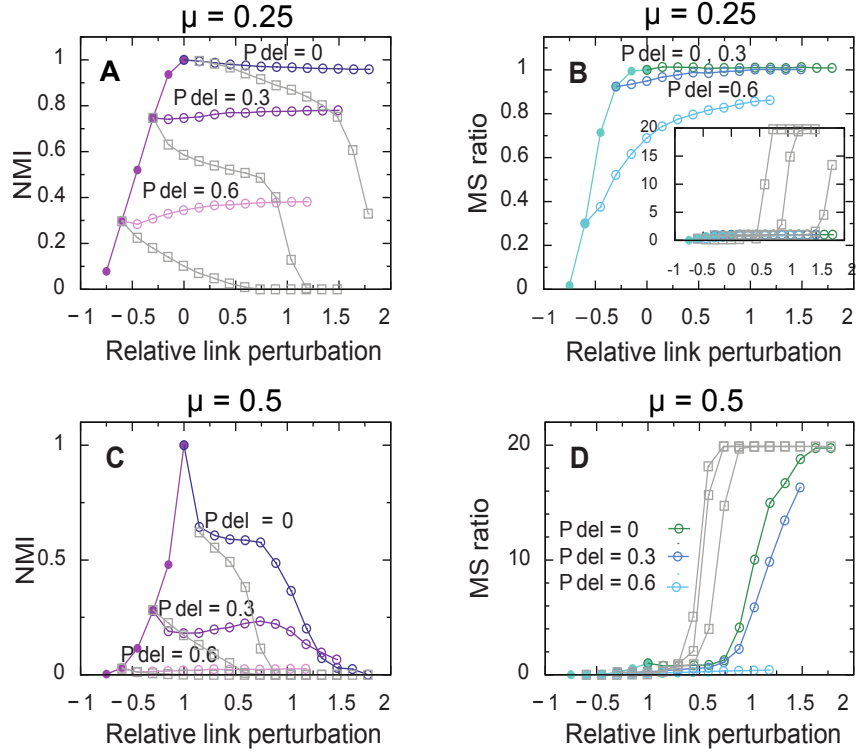}
\end{center}
\caption{ 
\textbf{Test of triangle completion on unweighted undirected benchmark networks.}
The panels show the similarity between the community structure of the original and the perturbed networks as a function of relative link perturbation, in A and B for low module mixing and in C and D for high module mixing. Panels A and C quantifies the similarity in terms of the normalized mutual information (NMI) and panels B and D quantifies the similarity in terms of the module size ratio. Filled circles correspond to the similarity after link removal. Open symbols correspond to the similarity after subsequently adding links by triangle completion (colored circles) and random link addition (gray squares). Link addition starts at 0, 30, and 60 percent link removal.
Each point corresponds to an average over 100 networks.}
\label{BnchTriCompCur} 
\end{figure}

Figure \ref{BnchTriCompCur} shows the result of using the perturbation method on benchmark networks with 1000 nodes, average degree $\langle k \rangle = 10$, community sizes between 10 and 50, and two different levels of mixing between communities. We generated sparse networks with missing links by randomly removing 30 and 60 percent of the links in the benchmark networks.
In the figure, we use \emph{relative link perturbation} to refer to the normalized difference between the number of links in the perturbed and unperturbed network.
The first row shows the result of triangle completion for low mixing, $\mu=0.25$, and well-defined communities. Low $\mu$, less than 0.5, means that, on average, each node has more links going to nodes within the same community than to nodes in other communities. So when we use our triangle completion method for perturbing such networks, we strengthen the structure inside the communities more than the structure between the communities. Therefore, we amplify the coarse-grain structure of the network, and the community structure of the perturbed network will be similar to the community structure of the original network, disregarding the number of extra links that we added. 
This reasoning is valid both when we perturb the original raw network and when we perturb the reduced networks. 
By adding extra links to the reduced networks, shattered and weakly connected modules aggregate and module sizes grow. For reference, the gray lines in Figure \ref{BnchTriCompCur} show that if we randomly add links, we completely destroy the community structure of the network.

\begin{figure}[!ht]
\begin{center}
\includegraphics[width=0.7 \columnwidth]{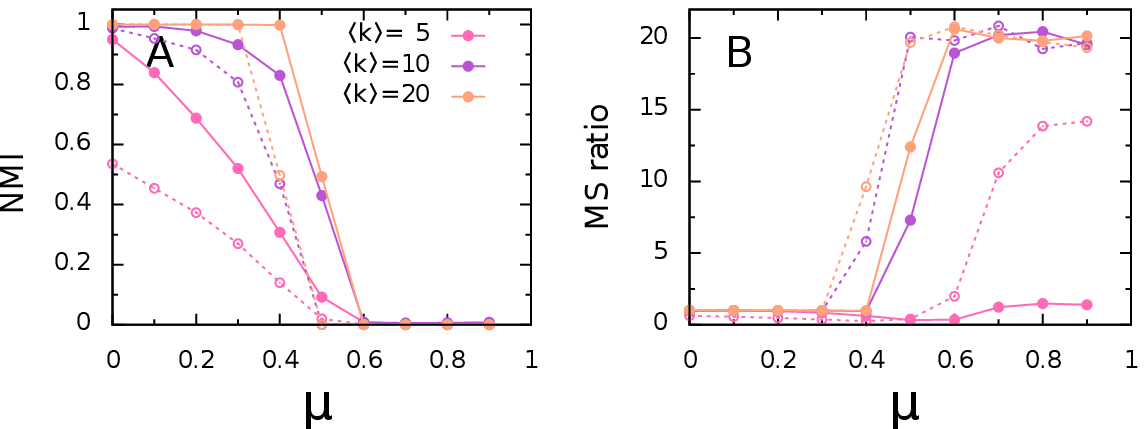}
\end{center}
\caption{
 \textbf{The success of triangle completion depends on the module mixing.} 
Similarity between the community structure of the original network and the perturbed networks for three different average degrees $\langle k \rangle$ as a function of the module mixing parameter $\mu$. In panel A the similarity is quantified in terms of the normalized mutual information (NMI) and in panel B the similarity is quantified in terms of the module size ratio. 
Filled lines and circles correspond to triangle completion and dashed lines and open circles correspond to random link addition.
No links were removed prior link addition and the number of links were doubled in all networks by link addition.
Each point corresponds to an average over 100 networks.}
\label{Mueffect}
\end{figure}

We use the ratio between the average module size of the perturbed network, $<S_i>_{link-added}$, and the average module size of the original network, $<S_i>_{original}$, to quantify module growth:
\begin{equation} \text{MS ratio}=\frac{<S_i>_{link-added}}{<S_i>_{original}} \end{equation} 
When the built-in community structure is well-defined for low $\mu$, the module size ratio does not exceed one and the community structure never collapses.
On the other hand, in networks with high $\mu$ and comparable number of links within and between communities, we destroy the community structure. That is, when we use triangle completion to perturb the network, module sizes grow quickly and finally collapse  (Figure \ref{BnchTriCompCur}C,D).
We have also analyzed networks with large communities, varying in size between 20 and 100 nodes, and found similar results. When completing triangles, the mutual information remains approximately constant and module sizes grow toward the original sizes as long as the mixing parameter is sufficiently low.
%On the other hand, in the networks with more complicated community structures ($\mu\leq0.5$), completing triangles destroy the community structures and module sizes grow quickly such that finally collapse happen.(Figure \ref{BnchTriCompCur} (c),(d))
In general, we find that $\mu=0.5$ is the threshold at which triangle completion works (Figure \ref{Mueffect}).
When $\mu$ is higher than 0.5, there are not enough regularities in the network to use for non-destructive perturbation.
Figure \ref{Mueffect} also shows that for denser and less challenging networks, the difference between triangle completion and random link addition decreases. For sufficiently dense networks, other methods, including link removal, can be used. But the more sparse the network is, the better is triangle completion over random link addition.

By repeatedly completing triangles and clustering link-added networks, we can generate bootstrap resamples for assessing significant communities in sparse networks with missing links. In the next section, we use this resampling technique to identify significant and insignificant communities in the network of ECJ case law.

\subsection*{ECJ case law network}
Case law is continuously evolving and changing over time. New cases build on old cases and areas of law emerge, vanish, evolve or remain constant over time.  Citation patterns between cases allow us to track and capture the evolution of areas of law. For example, Bommarito II et al.\ used a dynamic citation network to find meaningful clusters in the network of the United Supreme Court by means of a distance measure \cite{BommaritoII20104201}.
Here we use approximately 32,000 citations between more than 8,000 court cases (1954--2010) from the Court of Justice of the EU to better understand the overall structure of ECJ case law.

The European Court of Justice ensures the correct interpretation and application of EU law \cite{treaties19}.
When it comes to the judgments of the ECJ, legal scholars traditionally begin by distinguishing cases primarily concerning substantive issues from cases primarily concerning constitutional issues. Substantive issues regard questions about specific rights and obligations of individuals, Member States, and EU institutions under EU law. However, constitutional issues regard questions about the division of power between EU and Member States or the duties of Member States to enforce substantive rights. We find that the distinction between substantive and constitutional issues is supported by the network of ECJ case law. In addition to being substantive or constitutional, every judgment has also a procedural dimension in the sense that the ECJ enjoys jurisdiction over each case on one of eleven possible grounds \cite{treaties258-273}. More information about the Court's cases is available on the EU law website \cite{eur-lex}.

\begin{figure}[!ht]
\begin{center}
\includegraphics[width=0.4 \columnwidth]{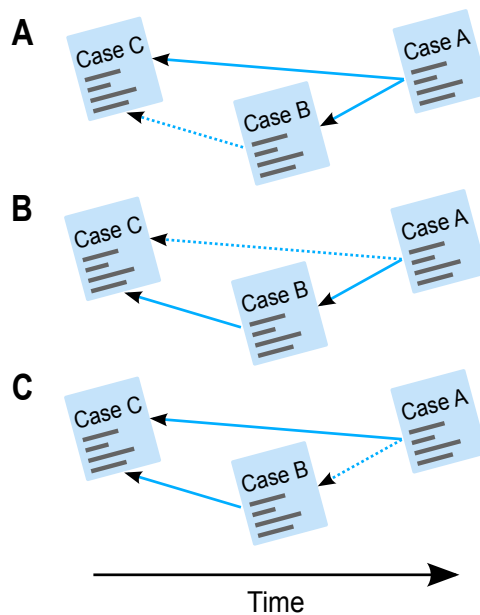}
\end{center}
\caption{
\textbf{Three possibilities for completing triangles in the time-directed network of ECJ case law.}
Given two citations between three cases, \textit{A} being more recent than \textit{B}, which in turn is more recent than \textit{C}, we can complete triangles in three different situations.
Panel A: If a new case \textit{A} cites two older cases \textit{B} and \textit{C}, but \textit{B} does not cite \textit{C}, we can make \textit{B} cite \textit{C}.
Panel B: If a new case \textit{A} cites \textit{B}, and \textit{B} cites \textit{C} but \textit{A} does not cite \textit{C}, we can make \textit{A} cite \textit{C}.
Panel C: If two new cases \textit{A} and \textit{B} both cite an old case \textit{C} and the newest case \textit{A} does not cite \textit{B}, we can make \textit{A} cite \textit{B}.
}
\label{TDbnch}
\end{figure}

We generated and clustered bootstrap networks from the network of ECJ case law to detect significant areas of law and to better understand the overall structure.
In the time-directed network of ECJ case law, each vertex corresponds to a court case and an arc from case \textit{A} to case \textit{B} shows that the newer case \textit{A} cites the older case \textit{B}, as schematically illustrated in Figure \ref{TDbnch}. Similar to many other time-directed networks, the network of ECJ case law is sparse, as, in the beginning, there were few cases to cite. However, because the number of cases increases with time, new cases have more options to cite.
Completing the triangles in the time-directed network of ECJ case law corresponds to one of the three situations depicted in Figure \ref{TDbnch}.
In all three situations, the added citation corresponds to a potential citation that we predict could have been considered and materialized in the first place.

\begin{figure}[!ht]
\begin{center}
\includegraphics[width=0.7 \columnwidth]{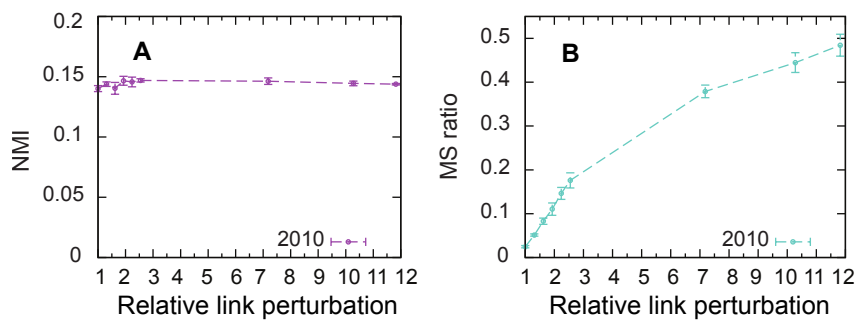}
\end{center}
\caption{
\textbf{Completing triangles in the court case network generates non-destroyed resample networks.} Panel A: Normalized mutual information (NMI) between the original network and the link-added networks as a function of the relative link perturbation. Panel B: Module size ratio between the original network and the link-added networks as a function of the relative link perturbation. Each point corresponds to an average over 100 runs.}
\label{CourtTriCompCur} 
\end{figure}

To show that our perturbation method does not destroy the core structure of the law network, we would like to compare the community structure of the link-added network to the community structure of the original raw network in terms of NMI.  
But the actual community structure of the original raw network is not known in this case. To overcome this problem, we use the case law directory code, the official classification system of the court, as our reference point. 
With this reference point, the NMI will be low but when we complete triangles we can use the trend of the NMI to validate our method.
As Figure \ref{CourtTriCompCur} shows, perturbing the ECJ case law network by completing triangles does not destroy the core structure of the network. For example, even when we make the network 12 times denser, NMI stays almost constant, but at the same time, the module sizes grow as we desire. 

For a significance analysis of the ECJ case law network, we first partition the network with a clustering algorithm to capture regularities in the raw network. To cluster with respect to citation flow between the court cases, we use the map equation framework with a generalized flow model for time-directed networks \cite{RosvallBergstrom08}. However, we emphasize that the significance analysis method works for any clustering algorithm. To assess the significance of detected clusters, we generate 100 resample networks by the triangle completion method without making any assumption about the underlying distribution of the resampled networks. Each resample network has twice the number of links as the raw network. Then we partition all resampled networks by using the same clustering method we used for the raw network. To identify significant clusters,  cluster cores, we search for the biggest subset of nodes in each cluster that gathered together in more than 90\% of the resampled networks. We define the size of a subset to correspond to the number of nodes in the subset and also to the volume of flow through the subset, weighted equally.
So by finding the core of each cluster, we can assess which nodes significantly belong to a cluster and which do not. In addition to identifying significant and insignificant nodes within each cluster, the resampled networks can provide us with information about which clusters are significantly stand-alone and which are probably subsets of other clusters. We consider a cluster as significantly stand-alone if its core is not partitioned with another cluster in at least 90\% of the resampled networks.
That is, two clusters are mutually insignificant if their cores are partitioned together in more than 10\% of the resampled networks. In this regard, each cluster could be insignificant with more than one other cluster, which means there is not enough support from the data for these clusters to exist as significantly stand-alone.
 
\begin{figure}[!ht]
\begin{center}
\includegraphics[width=1.1 \columnwidth]{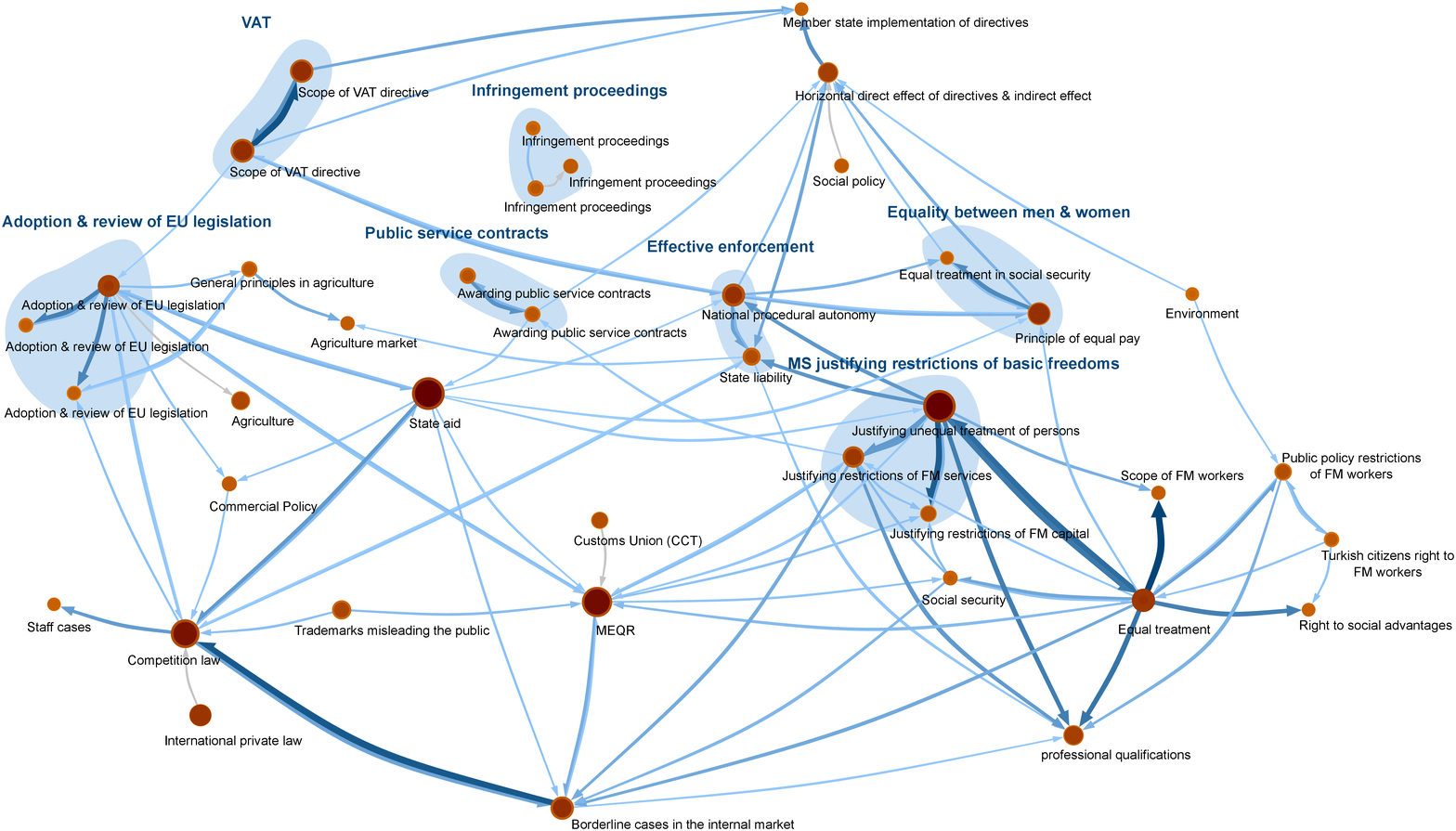} 
\end{center}
\caption{
\textbf{Map of ECJ case law.} We partitioned 8,200 court case documents with 32,000 citations. Afterwards, we generated 100 resampled networks using the triangle completion method. By clustering these resampled networks and comparing them to the clustering of the raw network, we can estimate how much support the data provide in partitioning the raw network. The map represents the 40 top modules. Insignificant clusters and their mutually insignificant friends are shaded with blue areas. }
\label{EU_map} 
\end{figure}

Figure \ref{EU_map} shows the map of the ECJ case law network illustrating the 40 top clusters, which we have manually named by analyzing which cases are clustered together. The size of nodes and links represent the citation flow within and between clusters, and we have highlighted mutually insignificant clusters by blue shaded areas. 

Several of the identified clusters represent well-established areas of law. One example is \emph{Equal treatment} (125 cases with 25 cases in the significant core, or 25/125 cases for short), which aggregates cases concerning discrimination of individuals based on nationality. Less intuitive, but seemingly valid, is the clustering of cases concerning the justification of such discrimination into a separate cluster, \emph{Justifying unequal treatment of persons} (113/134 cases). Interestingly, completing triangles aggregates not these two clusters but the latter with cases concerning Members States' (MS) justification of other violations of substantive rights in the highlighted area \emph{MS justifying restrictions of basic freedoms} in Figure \ref{EU_map}. Legal scholars have speculated in a convergence of these areas of law without being able to conclusively demonstrate this trend.
Another example of a structure that does not fit squarely into the traditional legal classification is \emph{Borderline cases in the internal market} (36/74 cases). The cluster works as a hub between different areas of law, bringing together cases involving several different substantive issues, including inter alia equal treatment.

The significance map in Figure \ref{EU_map} demonstrates that a single clustering of the sparse network is insufficient and can be misleading. For example, the map contains two clusters representing cases concerning Value Added Tax (VAT) (83/113 and 89/101 cases, respectively), even though there are no considerable differences between cases belonging to the two clusters. The significance analysis reveals that the two clusters are not significantly stand alone, because the significant cores are clustered together in 80 percent of all bootstrap networks. By completing triangles and aggregating the clusters, we can resolve the problem caused by missing links. The same is true for \emph{Public service contracts} (33/60 and 25/46 cases with 83 percent co-clustering of significant cores). The same is also true for \emph{Infringement proceedings} (10/58, 34/44, and 2/51 cases with 31 percent co-clustering between the least co-clustered pair of significant cores) and \emph{Adoption $\&$ review of EU legislation} (51/116, 31/43, and 5/38 cases with 31 percent co-clustering between the least co-clustered pair of significant cores). These clusters are also interesting because the cases are clustered based on the grounds for jurisdiction (procedural clusters), which would likely be absent in a more traditional legal categorization of the case law.

We also find, somewhat surprising from a legal perspective, that substantive, constitutional, and procedural clusters are closely related. For example, we find that there is a strong relationship between \emph{National procedural autonomy} (28/77 cases), which aggregates cases concerning the constitutional issue of procedural adequacy of national courts enforcing EU law, and \emph{The principle of equal pay} (74/84 cases), a cluster representing the substantive issue of the right of men and women to equal pay for equal work. The pattern of interconnected substantive and constitutional clusters remains on the level of aggregated clusters. Completing triangles and aggregating mutually insignificant clusters reveal a strong relationship between the highlighted constitutional area \emph{Effective enforcement} and the highlighted substantive area \emph{Equality between men and women}.

These results confirm that combining our resampling method with the significance analysis of the preliminary clusters can provide reliable aggregated clusters that help us better understand the modular organization of a system with missing information.

To summarize, using communities as the principal component of complex systems is reliable only if the communities are statistically significant and not the result of noisy or incomplete data. 
To assess the significance of communities in networks with missing links, we have suggested a simple approach that perturbs the sparse networks in a constructive way by adding links based on triangle completion. The remaining challenge is to estimate the optimal number of links to be added, but our benchmark tests indicate that results are insensitive to the number of added links. We used our method to identify significantly stand-alone communities and aggregate mutually insignificant communities in the sparse network of European Court of Justice case law. With a significance map of ECJ case law, for the first time we can analyze the large-scale organization of European law. We have, for example, identified structures and relationships that do not fit into the traditional legal classification system and empirically confirmed trends that legal scholars have only speculated in.

% You may title this section "Methods" or "Models". 
% "Models" is not a valid title for PLoS ONE authors. However, PLoS ONE
% authors may use "Analysis" 
%\section*{Methods}

% Do NOT remove this, even if you are not including acknowledgments
\section*{Acknowledgments}
We are grateful to Sara de Luna and Deborah Kolp for many valuable discussions. 
MR was supported by Swedish Research Council grant 2009-5344.

%\section*{References}
% The bibtex filename

%\bibliography{Court_2}

%\section*{Figure Legends}

%\section*{Tables}
%\begin{table}[!ht]
%\caption{
%\bf{Table title}}
%\begin{tabular}{|c|c|c|}
%table information
%\end{tabular}
%\begin{flushleft}Table caption
%\end{flushleft}
%\label{tab:label}
% \end{table}

\end{document}